\documentclass{www2003-submission}

\usepackage{rotating} 
\usepackage{a4wide}
\usepackage{latexsym}
\usepackage{graphicx}
\usepackage{color}
\usepackage{hyperref}
\usepackage[htt]{hyphenat}
\usepackage{makeidx} 
\usepackage{epsfig}

\newtheorem{algorithm}{Algorithm}

\begin{document}         

\title{The Best Trail Algorithm for Assisted \\ Navigation of Web Sites}
\numberofauthors{2}

\author{
  \alignauthor Richard Wheeldon\\
       \affaddr{School of Computer Science and Information Systems}\\
       \affaddr{Birkbeck University of London}\\
       \affaddr{Malet St, London}\\
       \affaddr{WC1E 7HX, United Kingdom}\\
       \email{richard@dcs.bbk.ac.uk}
  \alignauthor Mark Levene\\
       \affaddr{School of Computer Science and Information Systems}\\
       \affaddr{Birkbeck University of London}\\
       \affaddr{Malet St, London}\\
       \affaddr{WC1E 7HX, United Kingdom}\\
       \email{mark@dcs.bbk.ac.uk}
}

\balancecolumns

\date{\today}

\maketitle 

\begin{abstract}
We present an algorithm called the \emph{Best Trail Algorithm}, which helps solve
the hypertext navigation problem by automating the construction of memex-like
trails through the corpus. The algorithm performs a probabilistic best-first
expansion of a set of {\em navigation trees} to find relevant and compact trails.
We describe the implementation of the algorithm, scoring methods for trails, filtering
algorithms and a new metric called \emph{potential gain} which measures the
potential of a page for future navigation opportunities.
\end{abstract}

\section{Introduction}

The World Wide Web is a massive global hypertext system in which documents (or
pages) can be found on almost every subject imaginable. These pages are made
available by many authors and written in many languages. We consider a \emph{web
site} to represent a collection of pages with some common element, such as topic,
author or institution. The process of \emph{navigation} or \emph{surfing} is that
of following links according to the topology of the web site and viewing (or
\emph{browsing}) the contents of visited pages. During the navigation process users may become
``lost in hyperspace'', meaning that they become disoriented \cite{NIEL90}.
This happens when users fail to understand the context of the pages they are
viewing, are unsure of how they reached a page, cannot see how the page is
related to key pages such as the homepage or are uncertain as to where they
should proceed to find the information they are looking for \cite{LEVE02a}.

Vannevar Bush envisaged a hypothetical machine called a {\em memex}
\cite{BUSH45} - a cabinet-like box into which the user could store documents
and images. A sequence of such documents could then be annotated and linked
together to form a {\em trail}. By continuing the process, Bush imagined that
future workers could build a ``web of trails''.

In Berners-Lee's Web, a trail or \emph{navigation path} is implicitly formed
as the result of a navigation session in which the user visits a
sequence of web pages. Previous research \cite{BORG00} has shown how the trails
which users follow can be extracted from log data.
Often the starting point for one of these trails is a page resulting from
a search request \cite{NIEL97}, yet existing site search engines will neither
consider the possibilities for future navigation when returning their result
nor present details of the paths users might follow.

It is our hypothesis that constructing trails or paths in a query-dependant
manner will provide contextual information that will reduce the effects of the
navigation problem and increase user-satisfaction during search tasks.
Our contribution is to describe a probabilistic best-first algorithm for
automating the discovery of memex-like trails from a set of starting points. We
describe metrics for evaluating trails, and introduce a new metric for
determining more effective starting points by evaluating the \emph{potential
gain} of future navigation from a given page.
Previous hypertext systems have featured the ability to manipulate trails
manually \cite{CONK87,SILL90,REIC99} or allowed the construction of trails
using pure IR metrics \cite{BERN90,GUIN92}. However, none of these
systems has allowed the \emph{automatic} construction of trails by the computer
in any way that takes account of hyperlinks.

The rest of this paper is organized as follows:
In section~\ref{sec:trails} we describe our system for computing trails -
selecting starting points using the potential gain metric, expanding the
trails using the Best Trail algorithm and filtering redundant information
from them with heuristic methods.
In section~\ref{sec:evaluation} we describe our preliminary efforts to
evaluate the utility of the {\em navigation engine} which uses these
trails to assist users \cite{LEVE01a}.
In section~\ref{sec:implementation} we describe our implementation of the
algorithm.
In section~\ref{sec:experiments} we describe experiments into the behaviour
and performance of this implementation.
We discuss related work in section~\ref{sec:related} and give our concluding
remarks and directions for future research in section~\ref{sec:conclusions}.

\section{Computing Trails}
\label{sec:trails}

In this section we outline our methodology for computing trails.
Trails are computed by selecting relevant starting points, expanding a navigation tree
from each node using the Best Trail algorithm before filtering and sorting the
resulting set of trails.

We view a web site as a hypertext system $H$ having two components:
a directed graph $G = (N, E)$, having finite sets of nodes and edges $N$ and $E$,
respectively, and a scoring function $\mu$ which is a function from $N$ to the
set of non-negative real numbers.
The directed graph $G$ defines the web site topology and is
referred to as the \emph{web graph}; the nodes in $N$ represent the web \emph{pages}
and the edges in $E$ represent hyperlinks (or simply \emph{links}) between 
\emph{anchor} and \emph{destination} nodes. Figure~\ref{fig:topology} shows an
example web graph, taken from the GraphViz web site\footnote{
	\href{http://www.research.att.com/sw/tools/graphviz/}{www.research.att.com/sw/tools/graphviz}
}, which we will use as a running example. The terms node, web page
and URL will be used interchangeably. We interpret the score, $\mu(m)$ of a web page
$m \in N$, as a measure of how relevant $m$ is with respect to a given query,
where the query is viewed as the goal of the navigation session. The
\emph{Best Trail} algorithm computes trails scored by a function of these page
scores.

\subsection{Selecting Starting Points}

Whilst simply expanding from relevant points is effective, we can do
better by considering future navigation opportunities in our starting point
selection. We have created a metric for finding good starting points
which we refer to as the \emph{potential gain} of a url. That is, the potential
for future navigation opportunities.
Defined as the sum for all depths of the product of the fraction of
trails to that depth, $d$ and the discounting function $f(d)$, it is
easily computed by an iterative algorithm or by a series of matrix
operations \cite{WHEE03d,LEVE03}.
For larger graphs, we can utilize similar techniques to those proposed for
the PageRank citation metric \cite{PAGE98,HAVE99,KAMV03}.
For our experiments, we compute potential gain using the reciprocal function, $f(x)=x^{-1}$.

When restricted to a maximum depth of traversal, $d_{max}$ the naive
algorithm takes time proportional to $O(d_{max} . |E|)$ and space
proportional to $O(|N|)$ to compute potential gain values for all nodes
in $G$ given that $G$ is sparse. In practice, after
a brief settling period, convergence to a set of potential gain values
occurs in a short space of time. Bucketed values for potential gain
follow a power-law distribution, as is found for PageRank and many other
web-related phenomena \cite{ADAM02}.

\subsection{The Best Trail Algorithm}

\begin{figure*}[htb]
 \begin{center}
	\psfig{figure=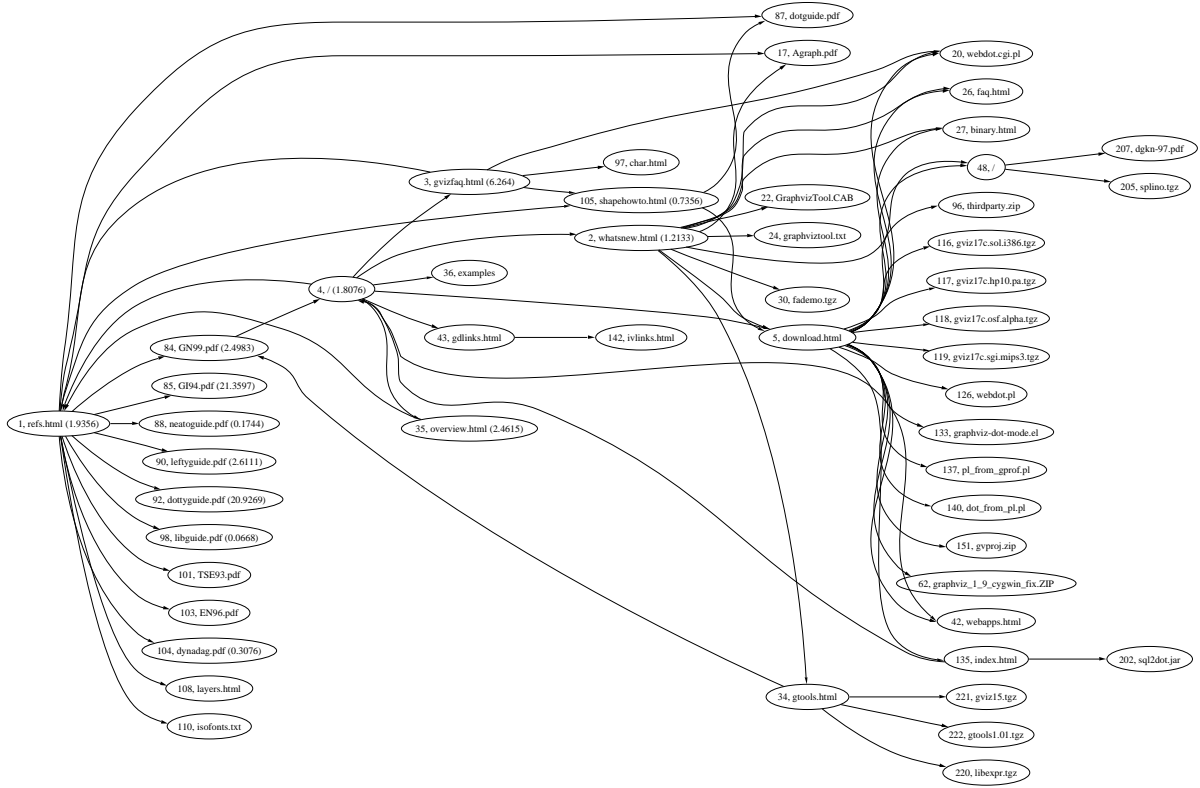, width=160mm}
	\caption{\label{fig:topology}
		An example Web topology, extracted from a crawl of the web site for the
		Graphviz project. The numbers denote unique Identifiers
		assigned to all URLs. The gaps in the sequence of IDs are due to URLs
		referenced by the website to pages elsewhere on the web. These URLs are
		reference, but the textual content of the pages is not indexed. The
		numbers in parentheses denote relevance scores for the query ``dotty''.
	}
\end{center}
\end{figure*}

The pseudo-code of the Best Trail algorithm is shown in figure~\ref{fig:best}.
It takes as input a set of starting URLs, $S$, and a parameter, $M \ge 1$,
which specifies the number of repetitions of the algorithm for each
input URL. When the algorithm terminates it outputs a set of trails, $B$.
There are $M$ trails in $B$ for each URL in $S$. Each trail is the highest
ranking trail contained within the \emph{navigation tree} expanded from a single
starting node. A navigation tree is a finite subtree of
the possibly infinite tree generated by traversing through $G$, the root of
which is a member of the set of starting points.
Manipulating sets of navigation trees has a filtering effect on the set of starting
points, reducing the rank of nodes which are isolated from other relevant documents
and from which navigation is problematic. Returning trails from separate trees
also has the effect of removing highly similar trails before further filtering
is required.

\begin{figure}[htb]
	\begin{algorithm}[Best\_Trail($S, M$)]\label{alg:best}
	\begin{rm}
	\begin{tabbing}
	t1\=t2\=t3\=t4\=t5\=t6\= \kill \\
	1.  \> \> {\bf begin} \\
	2.  \> \> \> {\bf foreach} $u \in S$ \\
	3.  \> \> \> \> {\bf for} $i = 0$ to $M$ {\bf do} \\
	4.  \> \> \> \> \> $D \leftarrow \{u\}$; \\
	5.  \> \> \> \> \> {\bf for} $j = 0$ to $I_{explore}$ {\bf do} \\
	6.  \> \> \> \> \> \> $t \leftarrow select(D)$; \\
	7.  \> \> \> \> \> \> $D \leftarrow expand(D, t)$; \\
	8.  \> \> \> \> \> {\bf end for} \\
	9.  \> \> \> \> \> {\bf for} $j = 0$ to $I_{converge}$ {\bf do} \\
	10. \> \> \> \> \> \> $t \leftarrow select(D, df, j)$; \\
	11. \> \> \> \> \> \> $D \leftarrow expand(D, t)$; \\
	12. \> \> \> \> \> {\bf end for} \\
	13. \> \> \> \> \> $B \leftarrow B \cup \{best(D)\}$ \\
	14. \> \> \> \> {\bf end for} \\
	15. \> \> \> {\bf end foreach} \\
	16. \> \> \> {\bf return} $B$ \\
	17. \> \> {\bf end.} \\
	\end{tabbing}
	\end{rm}
	\end{algorithm}
	\caption{
		\label{fig:best}
		The Best Trail Algorithm. The algorithm takes two
		arguments. $M$ is the number of repetitions and $S$ is a set of
		starting URLs.
	}
\end{figure}

\begin{figure*}[htb]
  \begin{center}
	\psfig{figure=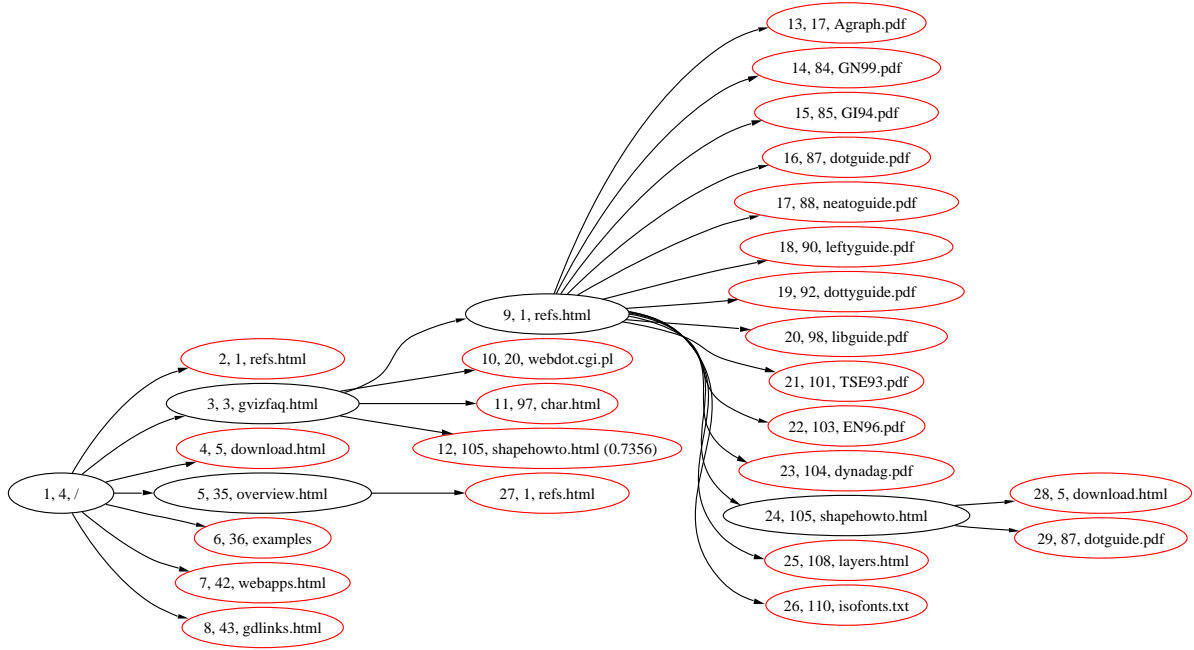, width=160mm}
	\caption[An example navigation tree.]{\label{fig:tree}
		An example navigation tree based upon the site structure shown in
		figure~\ref{fig:topology}. Each node is annotated with a unique tip id,
		a URLid, with the corresponding URL also shown. Red ellipses denote
		candidate tips for expansion. The tip numbers are assigned in sequence during
		the iteration of the algorithm. In this example, the tips numbered 1, 3, 9,
		5 and 24 were expanded. 
	}
  \end{center}
\end{figure*}

Starting from each node in $S$, the algorithm follows links from anchor
to destination according to the topology of the web site.
At each stage of the traversal, one of the $tips$ (the leaf
nodes of the navigation tree) is chosen for \emph{expansion}. The destination
node of each outlink whose source is represented by the chosen tip is assigned
a new tip which is added to the navigation tree, along with a computed
trail score. Previously visited nodes in the web graph will result in
distinct nodes in the navigation tree, with identical page scores but
different trail scores.
Figure~\ref{fig:tree} shows an example navigation tree based on the web
topology shown in figure~\ref{fig:topology}.

The algorithm has a main outer for loop which computes the best
trail for each URL. The second loop recomputes the best trail $M$ times.
The two innermost loops comprise the exploration and convergence stages of
the algorithm, both of which expand the navigation tree - from which the
best trail is selected by the $best()$ function. The number of iterations
in the exploration phase is set by $I_{explore}$, whilst the number of
iterations in the convergence phase is set by $I_{converge}$. During
the exploration phase, the $select()$ function selects a tip to expand
where the probability of a tip $t$ being selected is given by

\begin{displaymath}
	P(D_i, t) = \frac{\rho(t)}{\sum_{k=1}^{n} \rho(t_k)}
\end{displaymath}

where $\rho$ is a scoring function for the trail,
making the probability of any node being selected directly proportional
to its score. During the convergence phase, the probability of a node $t$ being
selected is dependant only on its relative rank, $\tau(t)$, in the ordered
set of candidate tips, and is given by

\begin{displaymath}
	P(D_i, t, df, j) = \frac{df^{\tau(t) j}}{\sum_{k=1}^{n} df^{\tau(t_k) j}}
\end{displaymath}

where $j$ is the number of completed convergence iterations and $0<df<1$ is a
\emph{discrimination factor}. The discrimination factor allows us to
discriminate between ``good'' trails and ``bad'' trails by reducing the
influence of trails with low scores. Thus during the convergence stage
``better'' trails get assigned exponentially higher probability. Setting $df$
equal to 1 would imply a uniform random selection, whilst as $df$ tends
towards 0, the behaviour of the algorithm tends towards that of a best-first
approach. The degenerate case of the Best Trail algorithm where $df=0$,
$I_{explore}=0$ and $I_{converge>0}$ is equivalent a simple best-first algorithm.
The \emph{rank} of a tip, $t$, (or of the trail leading to it), denoted
by $\tau(t)$, is determined by the tip's position within the ordered set
of candidate tips. The position of $t$ is determinated by comparing trails
based upon
\begin{enumerate}
	\item The number of query terms matched by the trail ending at $t$.
	\item The maximum number of query terms matched by any single page in
		the trail.
	\item The trail score, $\rho(t_k)$.
\end{enumerate}

It has been argued that the number of keywords in a query that are matched by
a document should take precedence over other scoring mechanisms, and that
the terms for a query may be spread across several pages \cite{ANH02,LI01,JOAC97}.
Ranking the trails first upon the number of keywords that are matched,
incorporates both of these ideas and improves relevance.

\subsection{Scoring Trails}
\label{sec:score}

We compute the relevance or {\em score} of a trail, $T = U_1, U_2, \ldots, U_n$,
as a function, $\rho$, of the scores of the individual web pages of the trail.
We need a function which encourages non-trivial trails whilst discouraging
redundant nodes. The following functions perform well in this regard:

\begin{enumerate}
\item The \emph{sum} of the scores of the \emph{distinct} URLs in the trail divided by the
	the number of pages in the trail plus some constant (e.g. 1).
  We refer to this scoring function as \emph{sum distinct}.
	This function penalises the trail when a URL is visited more than once. It
	also penalises trivial singleton trails and encourages trails where every
	node makes a significant contribution to the score.
	Removing the constant factor leads the objective function to return a
	maximal score in the case of a singleton node where that node is
	the highest scoring page in the corpus. Scoring functions such as the average score
	or maximum score of a node on a trail also suffer from this problem.

\item The \emph{discounted sum} of the  scores of the URLs in the trail,
	where the discounted score of $U_i$, the URL in the $i$th position in the 
	trail, is the score of $U_i$ with respect to the query
	multiplied by $\gamma$ and raised to the power of $i-1$,
	where $0 < \gamma < 1$ is the discount factor.

\item The \emph{weighted sum} of discounted scores, where the additional
	weighting is achieved by discounting each URL according to its previous number
	of occurrences within the trail. The weighted score of $T$ is given by
	\begin{displaymath}
		\rho(T) = weighted(T) = \sum_{i=1}^n \mu(U_i) \ \gamma^{i-1} \ \delta^{c(i)}
	\end{displaymath}
	where $c(i) = |\{ U_j|j<i \land U_j=U_i \}|$ and $\delta$ is a second discounting
  function, which reduces the importance of nodes with equal content.	We note that
	although	$i=j$ implies	$U_i=U_j$, $U_i=U_j$ does not imply $i=j$. Two distinct
	nodes may be considered equal if
	they have equal content, determined in advance using checksum of page contents
	and comparing likely candidates. This definition of node equality can easily be
	extended to refer to near-duplicate documents \cite{BROD00,SHIV99}.
\end{enumerate}

Figure~\ref{fig:scores} shows examples of score shows how the trails in the
navigation tree (figure~\ref{fig:tree}) would be scored after two expansions (of
tips 1 and 3). The examples shown in this paper are constructed by computing two
trails from each starting point - one scored using the {\em sum distinct} metric
and one using the {\em weighted sum}.

\begin{figure}[htb]
 	\begin{center}
     \begin{tabular}{|l|r|r|}
       \hline
       Tip & Weighted Sum & Sum Unique \\ \hline
       1   &       1.8076 &     0.9038 \\
       2   &       3.2593 &     1.2477 \\
       3   &       6.5056 &     2.6905 \\
       4   &       1.8076 &     0.6025 \\
       5   &       3.6534 &     1.4230 \\
       6   &       1.8076 &     0.6025 \\
       7   &       1.8076 &     0.6025 \\
       8   &       1.8076 &     0.6025 \\
       9   &       7.5940 &     2.5018 \\
       10  &       6.5056 &     2.0179 \\
       11  &       6.5056 &     2.0179 \\
       12  &       6.9194 &     2.2018 \\ \hline
     \end{tabular}
 		\caption { \label{fig:scores}
 			Table showing trail scores using Weighted Sum and Sum Unique.
			Example trails scores. The high score associated with the first trail has
			a useful control in forcing the most relevant pages to the forefront of
			the display. Merging trails with common roots gives a good ordering to
			the display, as can be seen in figure~\ref{fig:Dotty}
 		}
 	\end{center}
\end{figure}

\subsection{Sorting and Filtering}

The returned set of trails is unsorted and may contain redundant information.
To sort the trails would appear to be trivial - we simply apply the same rules
of sorting by number of keywords matches and then by the trail score. However,
we have more than one mechanism for scoring trails, and we can compute trails
in different navigation trees using different functions. We can sort the
resulting trails using a set of scoring functions, $F$, by specifying that a
trail, $T_1$ should be ranked higher than a trail $T_2$ if :
\begin{displaymath}
	\sum_{f \in F} \frac{f(T_1)}{f(T_1) + f(T_2)} >
	\sum_{f \in F} \frac{f(T_2)}{f(T_1) + f(T_2)}
\end{displaymath}

We can improve results by removing redundant trails and redundant sections
within trails. To achieve this, we need to define precisely what is meant
by a redundant trail.	We say that a trail $T_1$
subsumes a trail $T_2$ if and only if all the pages in $T_2$ are contained in $T_1$.
A trail, $t_1$ is removed from a result set, $r$ if and only if there exists a trail
$t_2 \in r$ such that $t_2$ subsumes $t_1$ and $\rho(t_2) > \rho(t_1)$.
Within a trail $T$, we consider a page, $t_i$ to be redundant if and only if the page
can be removed whilst still leaving a valid trail through the web site topology (i.e. if
$t_i$ is the last node of the trail or $(t_{i-1},t_{i+1}) \in E$ and the information
contained on page $t$ is either not relevant or contained in a previous page
(i.e. if $\rho(t)=0$ or $\exists j \ t_j=t_i \land j<i$). These definitions were
arrived at as the result of several experiments and typically remove
trivial reorderings and irrelevant content.

\begin{figure}[htb]
	\begin{center}
	\psfig{figure=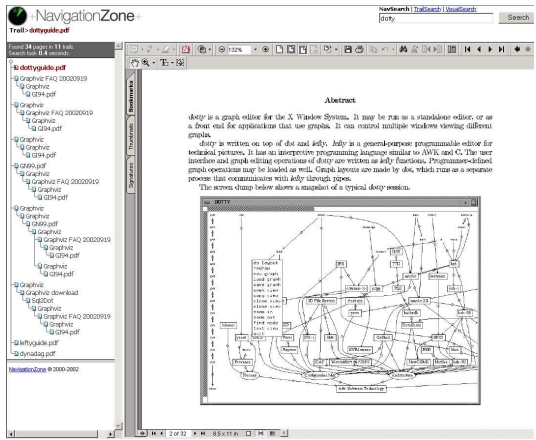,width=70mm}
	\caption{ \label{fig:Dotty}
		Screenshot showing the presentation of results for the query ``dotty'' on
		the topology shown in figure~\ref{fig:topology}.
	}
	\end{center}
\end{figure}

Finally, the trails with common roots are merged into a tree and presented in the
NavSearch UI \cite{LEVE01a}, shown in figure~\ref{fig:Dotty}. Two other interfaces have been

developed for displaying these trails - a flat TrailSearch interface similar
to that used by traditional search engines for displaying linear results and
a GraphSearch interface which displays the results in the form of a graph
\cite{WHEE02}

\section{Evaluation}
\label{sec:evaluation}

\subsection{A Case Study}

A case study was performed into the use of the navigation engine
on the Birkbeck School of Computer Science and Information Systems (SCSIS).
Queries were taken from a recent log file and
analysed. The chief results of the analysis are presented along with examples.

\begin{figure}[htb]
  \begin{center}
  \psfig{figure=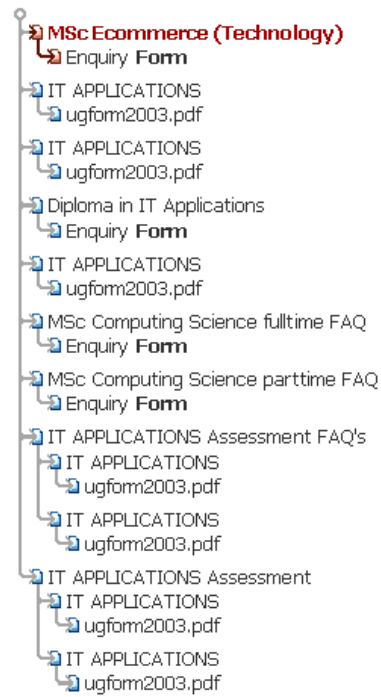,width=50mm}
  \caption{
		\label{fig:Dcs-ApplicationForm}
		Trails found for the query ``application form'' on the SCSIS site.
	}
	\end{center}
\end{figure}

\begin{figure}[htb]
  \begin{center}
  \psfig{figure=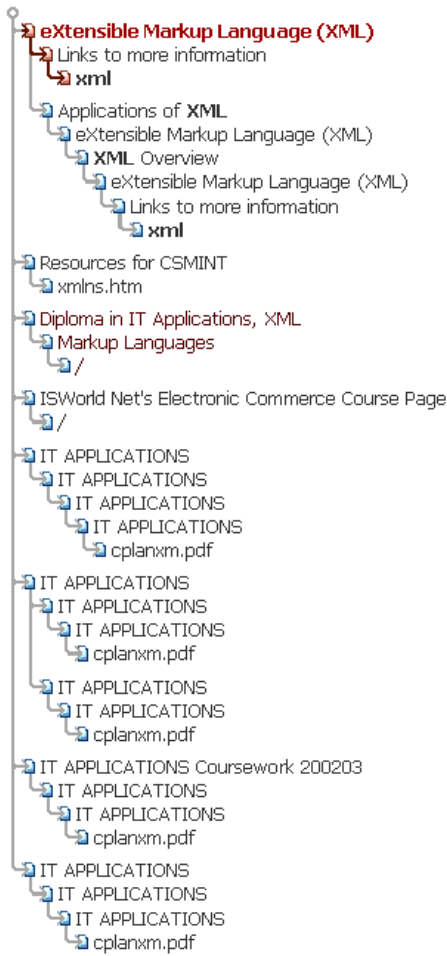,width=60mm}
  \caption{
		\label{fig:Dcs-xml}
		Trails found for the query ``xml'' on the SCSIS site.
	}
	\end{center}
\end{figure}

The trails provide relevant information. For example, results for the
query ``andrew'' find the home pages of Andrew Bielinski, Andrew Watkins and
Andrew Mair. For the query ``application form'', the first trail identifies
the application form for the MSc E-Commerce course and the second identifies
the application form required for the undergraduate program
(figure~\ref{fig:Dcs-ApplicationForm}). The first two trails for the query
``xml'', shown in figure~\ref{fig:Dcs-xml}, give brief tours of an
XML tutorial, always linking to external resources containing a great deal
of relevant information. The third trail provides an explanation of XML 
namespaces connected to hub with lots of XML references. The use of
Potential Gain in the starting point selection encourages such hubs to
be chosen. The fourth trail details the use and history of XML as a markup
language and it's relationship to SGML. Subsequent trails describe
the Information Technology (IT) applications module on XML.

\begin{figure}[htb]
  \begin{center}
  \psfig{figure=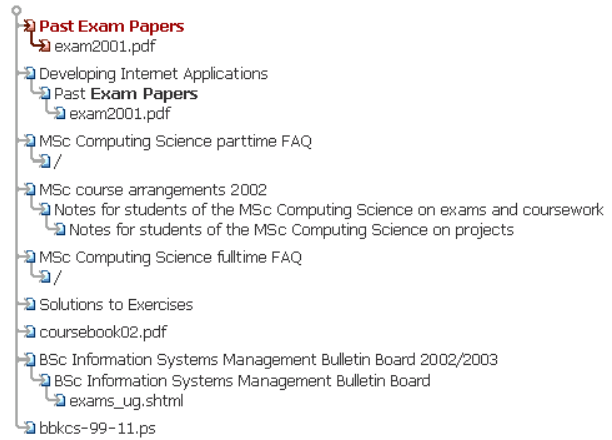,width=80mm}
  \caption{
		\label{fig:Dcs-ExamPapers}
		Trails found for the query ``exam papers'' on the SCSIS site.
	}
	\end{center}
\end{figure}

However, relevant content can be found with conventional, linear, search
engines. More important is that the trails provide {\em context} to show
associations and to help disambiguate the meaning of keywords and page
descriptions. For example, the structure of the trails for the query
``andrew'' shows Andrew Bielinski to be a research student under the
supervision of Mark Levene and that Andrew Mair is (although not a member
of the department) associated with the BSc Information Systems and
Management course. Similarly, for the query ``neural network'', the first
trail shows the course ``Artificial Intelligence \& Neural Networks'' linked
to the home page of Chris Christodoulou who teaches the course. Chris
Christodoulou is the SCSIS expert on nueral networks. The second trail leads
from his home page to the only one of his papers, ``A Spiking
Neuron Model: Applications and Learning'' linked to from his home page.
The user posing the query ``exam papers'' was almost certainly a student
looking for past papers for revision. Figure~\ref{fig:Dcs-ExamPapers} shows
that the first two trails provide exactly that. The second trail shows that
the papers relate to the module ``Developing Internet Applications''.
There are suprisingly few past papers available on the SCSIS
site and the remaining trails for this query details relating to arrangements
for sitting exams for that summer. The context provided by the trails makes
it easier to distinguish between the two types of result.

Unfortunately, the contextual information can be lost when inadequate short
titles are presented to describe the pages. For example, in
figure~\ref{fig:Dcs-xml}, it is impossible to tell any differences between the
page which share the title ``IT APPLICATIONS''.
Similarly, for the query ``accomodation'' (sic.), there are many different
pages shown in the trails, all of which relate to the Web
Dynamics workshop and contain the search term, but there is no means to
discriminate between them. The authors of the pages made no changes in the
{\tt h1} or {\tt title} tags by which to identify the differences. The most
appropriate title is contained in a later {\tt h3} tag.

The query ``accomodation'' also highlights another major problem - 
spelling errors are not corrected. Minor user errors or parsing errors in
the software introduce significant errors in the presented trails.
Similarly, examples such as ``birkbol programmes'', ``infirmation systems'' and
``Information Enginerring'' highlight the failure of users to
construct meaningful, accurate queries \cite{SILV99}.

Overall, the filtering operations appear to work well at reducing redundant
information without destroying contextual information. However, redundant
information appears commonly when near-duplicate documents cause separate,
highly similar, trails to be created. For example, in figure~\ref{fig:Dcs-ApplicationForm},
pages entitled ``IT APPLICATIONS'' are distinct but differ only by the inclusion
of an irrelevant ``assessment'' section. This small difference causes the creation
of 2 separate trails. This can be fixed with the application of near-duplicate
detection algorithms \cite{BROD00,SHIV99}.

The link structure can be broken when the crawler-based engine fails
to identify all the possible links. This can happen for several reasons - malformed
URLs, conservative robot exclusion policies \cite{KOST94}, javascript links and CGI
forms. For example, the link between
\href{http://eros.dcs.bbk.ac.uk/dept/rstudentperson.asp?name=bielinski}{rstudentperson.asp?name=bielinski}
and Andrew Bielinski's home page is missing, as are the links from all pages
in the SCSIS site to the home page,
\href{http://www.dcs.bbk.ac.uk/news/}{news},
\href{http://www.dcs.bbk.ac.uk/courses/}{courses},
\href{http://www.dcs.bbk.ac.uk/research/}{research} and
\href{http://www.dcs.bbk.ac.uk/seminars/}{seminars} pages.
Similar behaviour found with the output of Content Management
Systems (CMSs) such as Vignette or Documentum. The long-term solution
to this problem is to tie the trail engine into a better IR system and
offer interfaces to the main CMSs. For the current research prototype
this is not feasible, but would be essential if the navigation engine
was to be developed fully.

The conclusion that can be drawn from this analysis is that the trails found
by the navigation engine are useful, but the overall utility of the system is
being limited by problems with related modules - namely IR, near-duplicate
detection and short title generation. Given all these problem, the overall
performance of the system is highly promising. However, to truly test the
system's effectiveness requires an independent test with real users.

\subsection{A User Study}

\index{NavSearch}
In order to assess the usefulness of the NavSearch interface and prove the
hypothesis that ``a trail-based search and navigation engine improves users'
navigation efficiency'', Mat-Hassan and Levene conducted a usability
study. The results they obtained from the study revealed that
users of the navigation engine performed better in solving the question set
posed than users of a conventional search engine \cite{MATH01}.

Users were given two sets of information seeking tasks to complete based upon
the pages in UCL's official Web site. Three different search tools were evaluated,
one of which was the navigation engine with the NavSearch interface. The
others were Compass (UCL's official site search engine) and Google's university
search of UCL\footnote{ \href{www.google.com/univ/ucl}{www.google.com/univ/ucl} }.
Subjects were asked to answer two sets of questions, designed to be at the same
level of difficulty, using either NavSearch and Google or NavSearch and Compass.
The question sets were formulated so that all the questions fell within one of
five types : fact finding, judgement questions, comparison of fact, comparison
of judgement and general navigational questions.

Most of the subjects assigned to use Google were more optimistic about the
initial likelihood of completing the task, whilst those subjects assigned to
use NavSearch were initially more reserved and pessimistic. None of the subjects
had had any previous experience with NavSearch and familiarity was identified as
the main factor in favour of Google's linear interface model. Users were reported
to have ``found the interface quite intimidating'' considering it a ``radical
shift'' from the conventional layout and format of results.

The interfaces were assessed according to users' completion time, the number of
clicks employed, the number of correct answers found by the subjects and the
confidence and satisfaction levels expressed by the subjects. When asked to
compare NavSearch with Google or Compass, subjects expressed a much higher
degree of confidence in their ability to complete future tasks, a higher
degree of satisfaction with NavSearch with regards to the completion of tasks
and a higher degree of satisfaction completion with regard to navigation and
the display of results. Users stated that ``showing link relationship helps''
and that the system provided ``useful trails'' which gave ``an indication of
the pages already looked at and the pages that might be useful to look at''.
96\% of the study's subjects chose NavSearch over Google and Compass as their
preferred search engine. Mat-Hassan and Levene concluded that ``the proposed
user interface does indeed provide effective information retrieval assistance''.

\section{Implementation}
\label{sec:implementation}

In this section we give a brief outline of the architecture required to support
trail finding and details of the algorithm's implementation.

Each node, page or URL is assigned a unique ID. IDs are 32-bit signed integers
assigned in sequence (from 1) to each URL such that any two identical URLs will
have an identical ID. The mapping between URLs and IDs is performed using
Berkeley DB files \cite{SLEE01}. Each page is associated with a relevance score,
determined using $tf.idf$ measures
although they may be computed using any information retrieval
metric \cite{SALT98,BAEZ99}. Given a set of relevances and a graph in
this form, we compute the best trails by running the traversal stages in
a separate threads for each starting point.

There are many ways to access relevance data in constant time - either through
array lookups or hashtables, depending on the size of the webcase. The graph is
stored using the URL ids as references. Many strategies have been presented for
returning sets of inlinks and outlinks from large graphs with appropriate
time-space trade-offs \cite{BHAR98,RAND01,BOLD03}.

At each step of the expand and converge process we must select a tip for
expansion based upon the probability distribution described in
section~\ref{sec:trails}. These distributions have been carefully selected
to allow the use of binary trees for storing this trail score information.
We can implement this efficiently by using a table describing the tip selection
tree at each stage, reducing the object creation overhead. Associated with
each tip is the sum of all relevances for
all descendants, denoted as the subscore, $s$, and the total number of
descendants which are referred to as the subcount, $c$.
Figure~\ref{fig:candidates} shows the table storing the tips of
the navigation tree shown in figure~\ref{fig:tree}.
 
\begin{figure}[htb]
 	\begin{center}
		\small{
     \begin{tabular}{|c|c|c|c|c|c|c|}
       \hline
       Tip & Weighted Sum & Left & Right & SubScore & SubCount \\ \hline
       1   &       1.8076 &    2 &     4 &  49.9809 &       12 \\
       2   &       3.2593 &    3 &       &  40.9429 &        7 \\
       3   &       6.5056 &    9 &     5 &  37.6836 &        6 \\
       4   &       1.8076 &      &     6 &   7.2304 &        4 \\
       5   &       3.6534 &   10 &       &  16.6646 &        3 \\
       6   &       1.8076 &      &     7 &   5.4228 &        3 \\
       7   &       1.8076 &      &     8 &   3.6152 &        2 \\
       8   &       1.8076 &      &       &   1.8076 &        1 \\
       9   &       7.5940 &      &    12 &  14.5134 &        2 \\
       10  &       6.5056 &      &    11 &  13.0112 &        2 \\
       11  &       6.5056 &      &       &   6.5056 &        1 \\
       12  &       6.9194 &      &       &   6.9194 &        1 \\ \hline
     \end{tabular}
		}
 		\caption { \label{fig:candidates}
 			Table showing candidate tips for expansion. SS is the sum of the scores for
 			the current node and all descendants and SC is the number of active nodes
 			reachable from that node. It should be noted that the nodes in this tree represent
 			tips and should not be confused with either the nodes of the graph or the navigation
 			tree produced by the Best Trail.
 		}
 	\end{center}
\end{figure}

When selecting a tip to expand, a random number between 0 and $x$ is selected
where either $x$ is the subscore or
\begin{displaymath}
	x=\sum_{k=0}^{c-1} df^{\tau(t_k) j k}
\end{displaymath}
which can be computed in constant time by applying the known result for the
sum of a geometric series\footnote{ $\sum_{k=x}^{y} a^k = \frac{a^x (1-a^{y-x+1})}{(1-a)}$ }.
At each step in the subsequent traversal, this process is repeated for the
nodes to the left and right of the current node, adjusting $x$ and $y$
appropriately. Thus, the interval in which the selected value lies can be
chosen and a direction selected. Once completed a single tip will remain,
which is then expanded. For example, in an expansion iteration, the process
would start with the selection of a random number between 0 and 49.9809.
If the number 49 was chosen, the process would proceed to the right. If the
number 35 was chosen, the process would proceed to the left.

\subsection{Complexity}

It has been shown how the step $select(D_i, df, j)$ can be implemented
to run in time O($ \log (n)$) where $n$ is the number of candidate tips. The
function $best()$ has the same time complexity, but is slightly
simpler in that each iteration is to the left of the current node.
Hence, the worst case complexity of algorithm~\ref{alg:best} using this implementation
can be given as O($K M I^2 \beta^2$) where $I = I_{explore} + I_{converge}$
and $\beta$ is the maximal outdgree of any link in $E$. This can be broken down
as follows:
\begin{description}
  \item [$I \beta$] as the worst-case insertion time for a tip. This factor
		emanates from the fact that the tree of tips may become
		a linked list if all new tips are added to the same part of the tree.
		This might occur in the simple case of nodes having identical scores,
		so these scores are biased using tiny random numbers to adjust the rank.
		The magnitude of these adjustments means that they affect only the speed
		of the operation, not the end results.
  \item[$\beta$] representing the number of potential tips which may
		be added to the candidate set at each iteration. This number would
		always be added on a fully connected graph, but graphs based upon Web
		data are very sparse and this will never occur in practice.
	\item[$KMI$] as the maximum number of iterations the Best Trail may take
		to find the given trails.
\end{description}

In practice the tree of tips is unlikely to be skewed to such a degree. Nor
is the graph likely to be fully-connected. However, if the average-case
complexity is performed by substituting the average outdegree, the results
are still inaccurate. Using the {\em weighted average outdegree} better models
the expansion of the navigation tree during the expansion and convergence
phases.
The weighted outdegree, $W$, of a node, $n$, is defined as the
product of the number of outlinks $(n,x)$ from that node and the proportion
of links in the graph which point to that node $\frac{|{n,y \in E}|}{|E|}$.
It is assumed that all links are as likely to be followed as any others, given a
sufficient number of queries. It should be noted that, when expanding a
navigation tree, the number of potential trails to a depth of $d$ is roughly
equal to $\sum_{i=1}^{d} w^i$. where $w$ denotes the weighted average outdegree
of a graph. Given that $\beta$ is the weighted average outdegree, the average
case complexity can be given as O($K M I \beta \log (I \beta)$). Using binary
trees the average-case
complexity of the $expand$ operation is O($\beta \log \beta I$) since there are,
on average, $\beta$ elements to be added to the list of candidate tips and the
complexity of operation to insert these new candidates is equal to that of the
$select$ function - O($\log \beta I$).

\section{Experimental Results}
\label{sec:experiments}

We have conducted numerous experiments to test the behaviour of the algorithm
and explore the effect of the various parameters which control it. These were
mostly performed on crawls of the Birkbeck website, the school of computer
science and information systems website and the JDK 1.4 javadocs, primarily
due to the abundance of query information available to us.

\begin{figure*}[htb]
  \begin{center}
    \begin{tabular}{cc}
			\psfig{figure=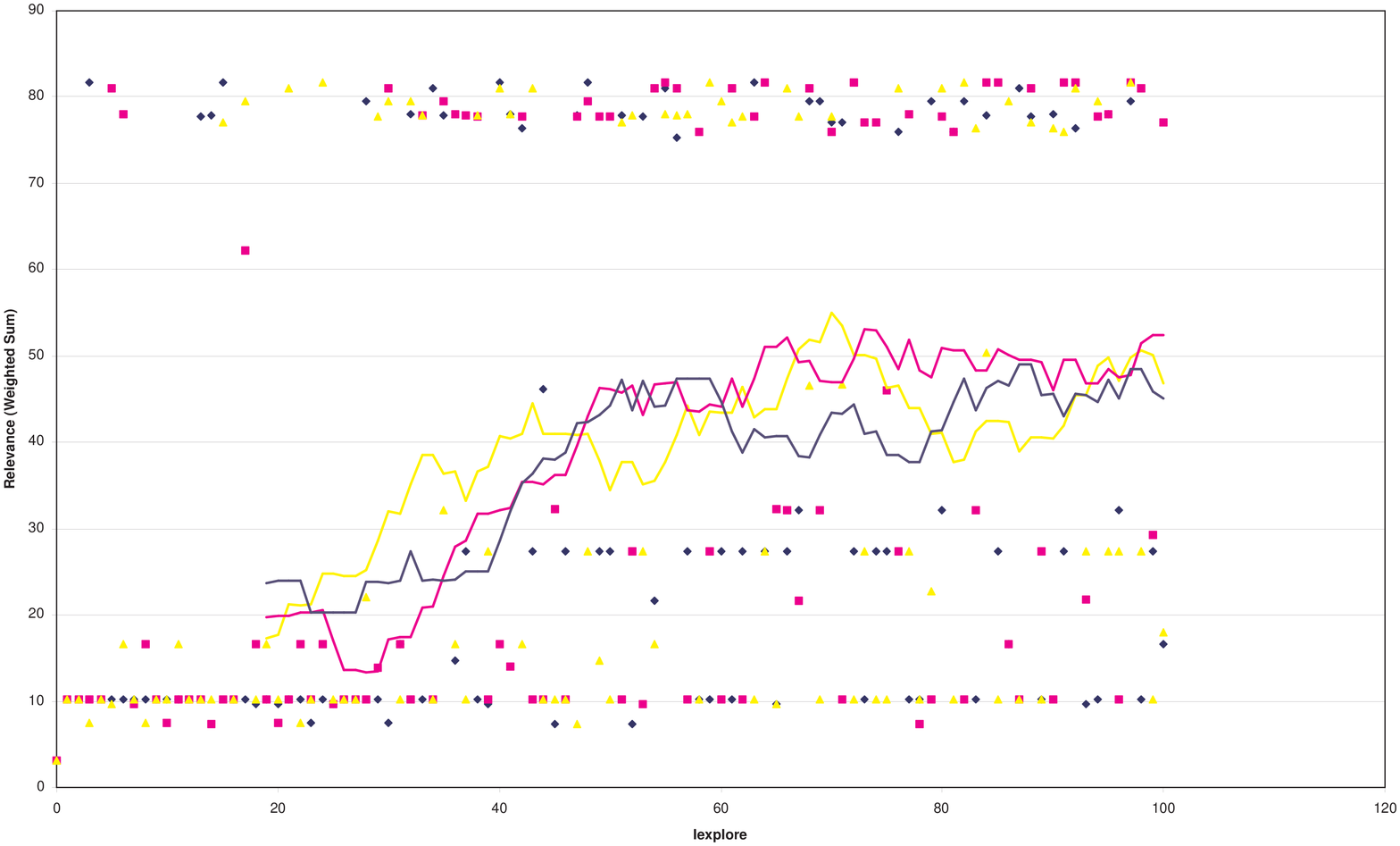,width=75mm} &
			\psfig{figure=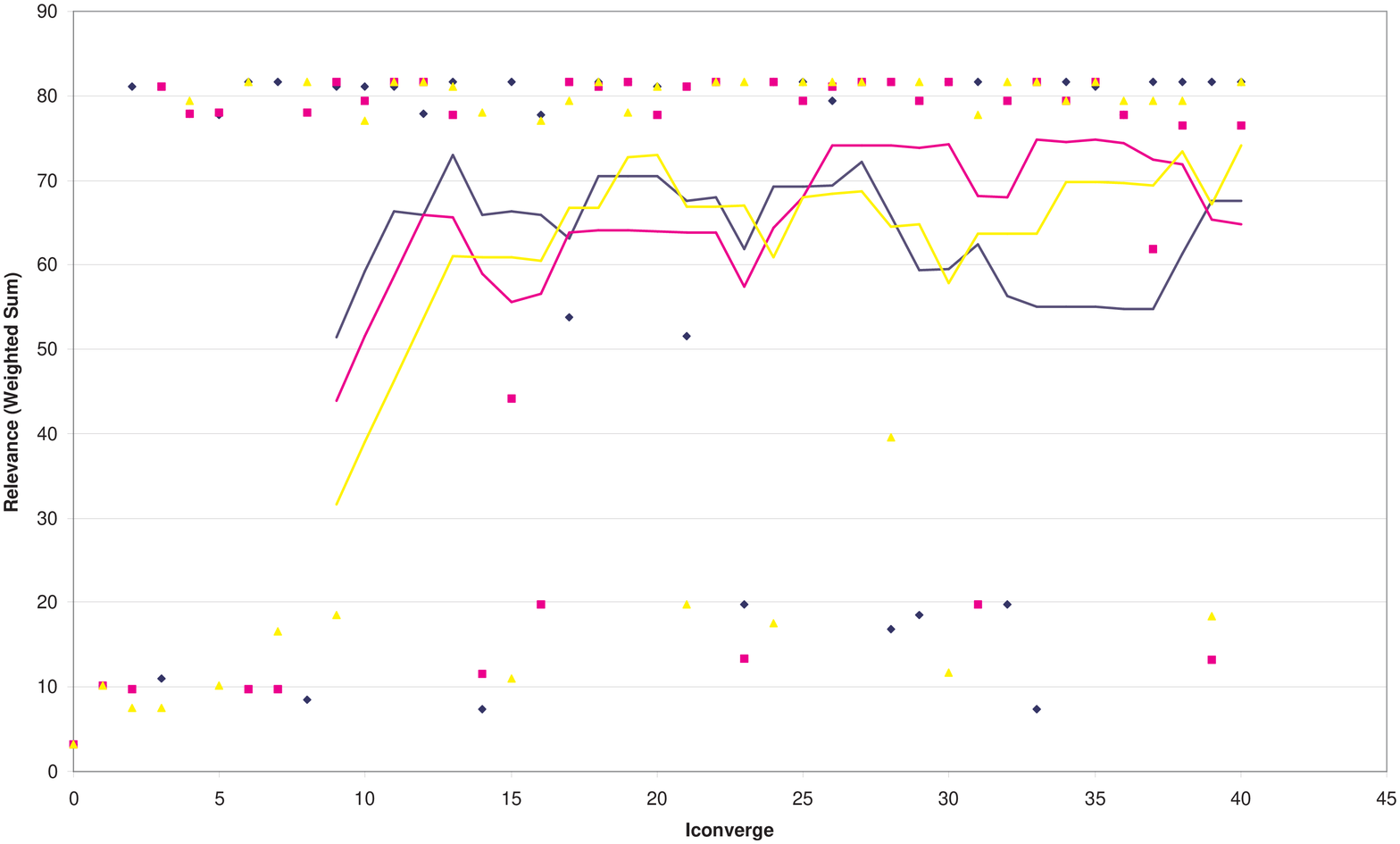,width=75mm} \\
      (a) $I_{explore}$ &
      (b) $I_{converge}$ \\
		\end{tabular}
	\end{center}
	\caption{ \label{fig:Swing}
		Increasing either (a) the number of exploration iterations or (b) the number
		of convergence iterations, increases the score of the returned trails. When
		increasing $I_{explore}$, the algorithm slowly tends to a limit, whilst
		exploring the solution space. When increasing $I_{converge}$ (and leaving
		$I_{explore}$ constant e.g. 0 as in this example), the algorithm quickly tends
		to a limit.
	}
\end{figure*}

\begin{figure*}[htb]
  \begin{center}
    \begin{tabular}{cc}
			\psfig{figure=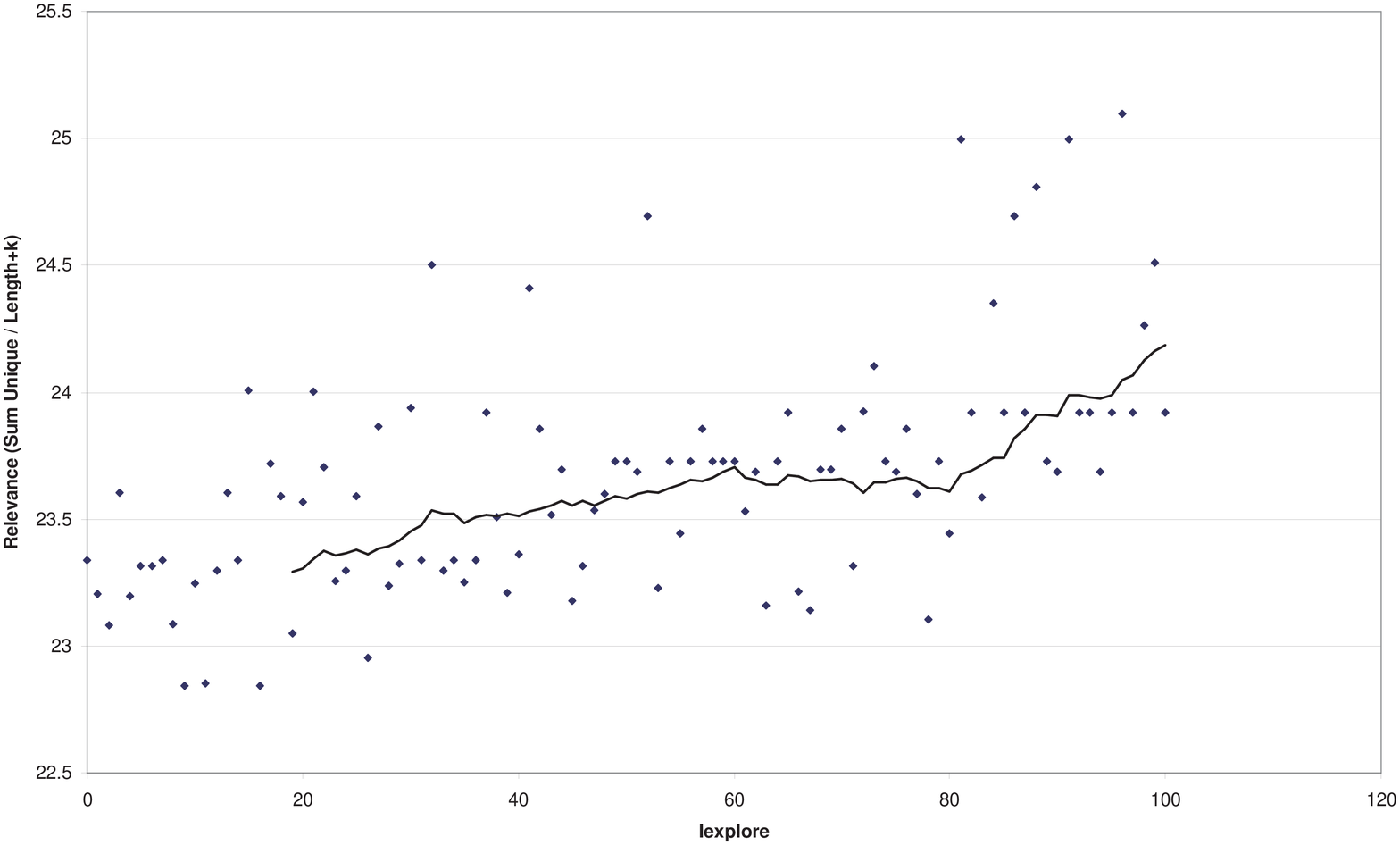,width=75mm} &
			\psfig{figure=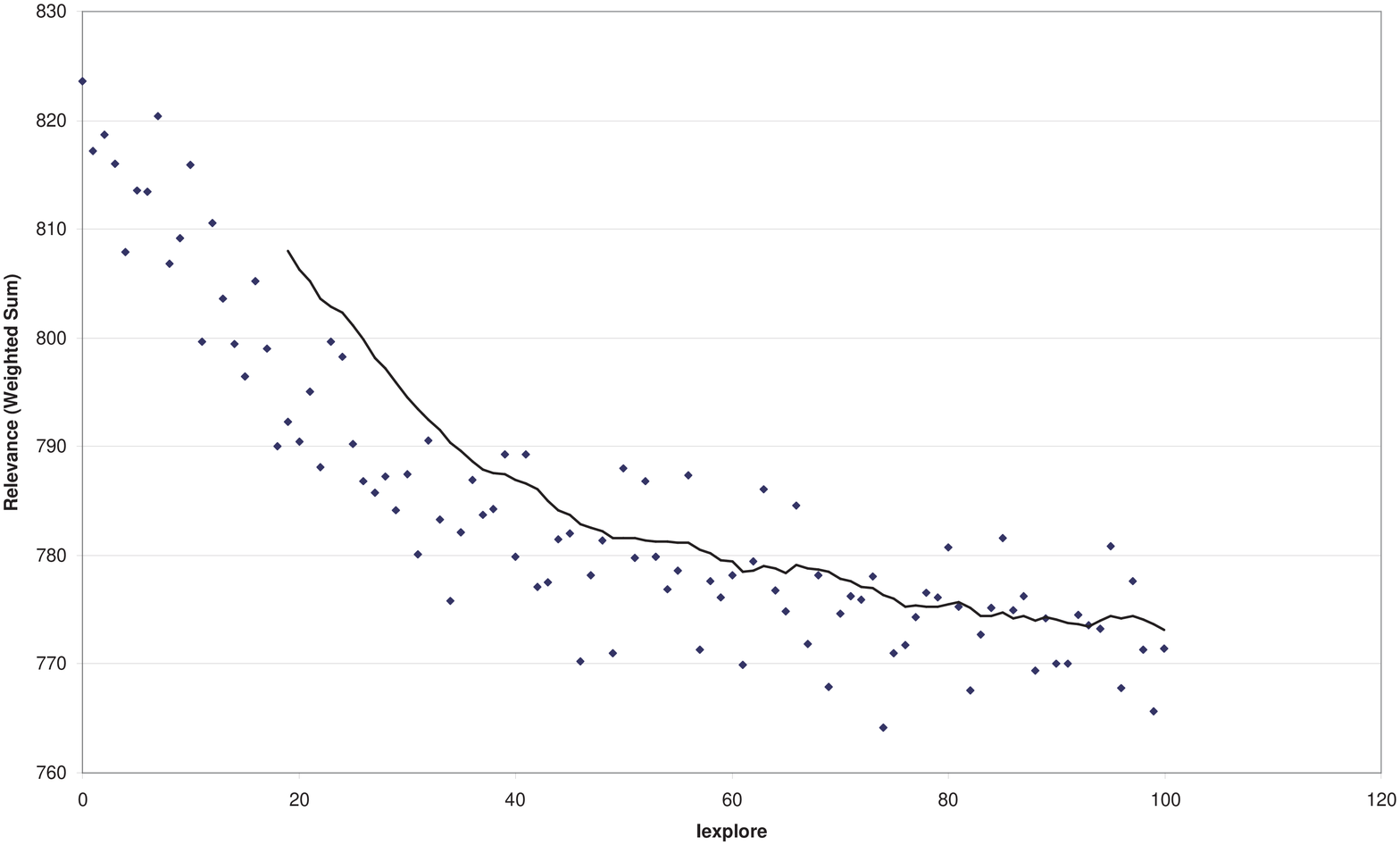,width=75mm} \\
      (a) Sum Distinct &
      (b) Weighted Sum \\
		\end{tabular}
	\end{center}
	\caption{ \label{fig:IexpandOrIconverge}
		Increasing Iexpand, whilst decreasing $I_{converge}$ increases the resultant
		trail score when calculated using \emph{sum distinct} but decreases the
		resultant trail score when calculated using the \emph{weighted sum}
		The graphs show values for $0 \le I_{explore} \le 100$ where $I_{converge} = 100 - I_{explore}$.
	}
\end{figure*}

\begin{figure}[htb]
	\psfig{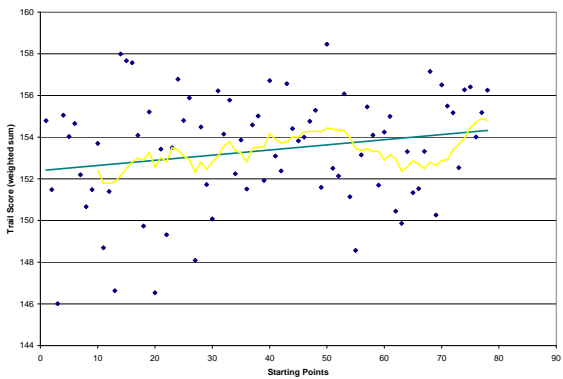}
	\caption{ \label{fig:AutodocStartingPoints}
		Increasing the number of starting points increases the score for trails, by
		allowing a greater number of opportunities for discovery. Trail sets are
		truncated to the same size.
	}
\end{figure}

Behaviour of the algorithm is controlled by the parameters $df$, $I_{explore}$,
$I_{converge}$, $M$ and the set of starting points $\{U_0, U_1, \ldots, U_K\}$.
As we would expect, increasing the value of either of the parameters $I_{explore}$
or $I_{converge}$ produces higher scoring trails on average (figure~\ref{fig:Swing}).
Unsuprisingly, increasing $I_{converge}$ finds the local limit of the trail score
faster than increasing $I_{explore}$, as shown by the sharp rise at the very start of
the curve. Perhaps more suprising is the behaviour when altering the ratio between
$I_{explore}$ and $I_{converge}$. Increasing $I_{explore}$ whilst decreasing $I_{converge}$
increases the scores of the resulting trails if we measure the relevance using
\emph{sum distinct} but decreases the trail score when calculated using the
\emph{weighted sum} (figure~\ref{fig:IexpandOrIconverge}).
The balance between the values $I_{explore}$
and $I_{converge}$ can be tuned to reflect the importance of the two metrics.
Increasing the value of $M$ is less effective, as repeated exploration from
the same node causes many of the expansions to be duplicates of those performed
in other trees. We can use the multi-treaded environment better by expanding from
a greater number of starting points, as shown in figure~\ref{fig:AutodocStartingPoints}.

In order to evaluate the effectiveness of the Potential Gain metric in improving
trail scores, we analysed the scores of trails found by traversing the graph from
starting points selected by combining the $tf.idf$ IR measure, $\mu(p)$, of a page
$p$ with the page's potential gain, $Pg(p)$ in several different ways. Comparisons
were also made to test the effectiveness of a simple outdegree count, $Out(p)$ and
of Kleinberg's hub metric\cite{KLEI98}. The results showed that, relative to the
baseline of selecting according to $\mu(p)$, a significant improvement is achieved
by taking the highest scoring pages when scored using $\mu(p) Pg(p)$ or
$\mu(p) \log Pg(p)$. Suprisingly, the simple metric $\mu(p) \log Out(p)$ also
performed well for the task of starting point selected whilst Kleinberg's metric
performed badly.

\section{Related Work}
\label{sec:related}

Many graph traversal and path-finding algorithms have been developed over the
last 50 years and it is not unreasonable to question the development of a new one.
We will consider the effects of a few of them. A depth-first traversal, for example
is unsuitable for trail finding as it may tend towards ``black-holes'' from which
there is no escape. It is considered unsuitable for crawling for similar reasons.
Breadth-first search is non-viable for anything other than very short trails,
due to the exponential growth of the tree. A best-first search is possible but
will struggle in situations where the best pages are separated by content which
is less relevant - exactly the situations where automated navigation is most needed!
Another approach that has been used effectively for computing solutions to the
Travelling Salesman Problem (TCP) is Ant Colony Optimization (ACO) \cite{DORI96}.
Each ``ant'' is an agent which uses a greedy heuristic to follow a trail based upon
the weight of links and the presence of a ``pheromone''. This pheromone is laid by
ants following a path, based upon the length of the final result. Our own experiments
have provided anecdotal evidence that the Best Trail algorithm out-performs ACO for
web-site trail finding, although the ACO system appears to out-perform the Best Trail
in finding solutions to TSP.

Several systems have allowed the manual construction of trails. Sillitoe et al.
\cite{SILL90} proposed a system for manipulating trails, complete with forks and
subtrails. They discussed a database backed scheme for storing and retrieving the
information. Furuta et al. \cite{FURU97} developed a system for authoring
modifying and re-using \emph{Walden's paths} - guided tours which could be used
in a teaching environment.
WebWatcher advises users on navigation possibilities by highlighting links as
they browse. This forms a trail over time, but the link-at-a-time approach does
not allow the user to see the context initially. We agree with Joachims et al.'s
\cite{JOAC97} belief that ``in many cases only a sequence of pages and the knowledge
about how they relate to each other can satisfy the user's information need'',
but extend this to compute and show complete sequences in advance.
Bernstein's approach to constructing trails was to ask the user to ``choose an
interesting starting point and ask the apprentice to construct a path through
related material''. The tours were constructed via a best-first page finding
scheme using document similarity measures \cite{BERN90}.

The concept of Information Units, presented in \cite{LI01} also attempts to break
away from the single page model, returning small clusters of linked pages
answering the user's query. The returned units may be more compact than the
trails returned by the best trail, and cover situation which cannot be
handled using trails, but the returned results are not navigable, nor has there
been sufficient consideration to the display of the results and subsequent
user interaction.
The Cha-Cha system \cite{CHEN99} presents results in a similar manner to the
NavSearch interface and shows results in context, but the scoring is only
conducted at the page level, the trails leading to the page are chosen as
the shortest paths, not those with informative content.

Several metrics have been proposed for selecting nodes in search results
which relate to the issue of starting point selection. The most famous, the
PageRank citation \cite{PAGE98} only considers the effect of incoming links,
whilst Kleinberg's Hubs and Authorities metrics and extensions of it only
consider the effect of single links in each direction, whilst potential gain
will consider the effect of more distant pages \cite{KLEI98,LEMP00}.

\section{Conclusions and Future Work}
\label{sec:conclusions}

We have presented an algorithm for finding trails across the graph of linked pages
in a web site. Inspired by Bush's memex, these trails provide a structure to the
returned results and provide users with contextual information not provided by
traditional search facilities.

Although site-search is of vital importance, and deserves special attention as an area
of research separate from global search engines, it would be highly beneficial to allow
full web-scale trail finding. Unfortunately, the current architecture will
not scale to full-size web data. However, we can break the problem down.
Conventional search engines do not index the full content of the web. They select
some subset to index based on usage statistics, link analysis or the output of
dedicated crawling algorithms designed to select high-quality nodes first
\cite{PINK00,CHO98}. We can select a subset of this on which to perform
trail computation. For example, we could compute trail information on
high-profile or highly-popular sites and return single-page results for the remaining
indexed pages. An alternative strategy is to construct a restricted graph based upon
the search results for a given query, over which trails could be constructed. Whilst
this approach would suffer less scalability problems, it might suffer similar
performance issues to Kleinberg's approach of expanding the search results \cite{KLEI98}.

The work presented here has many applications in other, non-hypertext areas.
We have built a system called DbSurfer, which applies these ideas to solve the join
discovery problem in relational databases by finding trails through the graph of
foreign key dependencies.
We have also built systems for finding trails in program documentation \cite{WHEE02}
and source code. In this last example, the results are achieved by combining
trails discovered on several graphs, where each graph corresponds to interactions
in one of five different coupling types (Inheritance, Interface, Aggregation,
Parameter Type and Return Type) \cite{WHEE03c}. In these examples, the problems
identified earlier are largely eliminated and the true potential of the trail-based
navigation engine can be clearly seen.
The navigation problem is widespread and occurs in all type of software system. Alan Cooper
describes the phenomenon as ``uninformed consent'', when ``at each step
the user is required to make a choice, the scope and consequences of which are not
known'' \cite{COOP99}. Providing keyword search and trail discovery over the graph
of options available at the application or operating system level could greatly
enhance user experience. For example in Microsoft Windows, the query ``active desktop''
might return a path $\texttt{Start} \rightarrow \texttt{Settings} \rightarrow
\texttt{Control Panel} \rightarrow \texttt{Folder Options}$.
Finally, we believe that the algorithm may have applications in the fields of game
playing and optimization problems.

\bibliographystyle{plain}
\bibliography{../bib/papers}

\begin{thebibliography}{10}

\bibitem{ADAM02}
Lada~A. Adamic.
\newblock {\em Network Dynamics: The World Wide Web}.
\newblock PhD thesis, Stanford, 2002.

\bibitem{ANH02}
Vo~Ngoc Anh and Alistair Moffat.
\newblock Impact transformation: effective and efficient web retrieval.
\newblock In {\em Proceedings of the 25th annual international ACM SIGIR
  conference}, pages 3 -- 10, Tampere, Finland, 2002. ACM Press.

\bibitem{BAEZ99}
Ricardo {Baeza-Yates} and Berthier {Ribeiro-Neto}.
\newblock {\em Modern Information Retrieval}.
\newblock ACM Press and Addison Wesley, Reading, Ma., 1999.

\bibitem{BERN90}
Mark Bernstein.
\newblock An apprentice that discovers hypertext links.
\newblock In Antoine Rizk, Norbert~A. Streitz, and J.~Andr\'e, editors, {\em
  Hypertext : Concepts, Systems and Applications : Proceedings of the First
  European Conference on Hypertext}, The Cambridge Series on Electronic
  Publishing. Cambridge University Press, November 1990.

\bibitem{BHAR98}
Krishna Bharat, Andrei Broder, Monika Henzinger, Puneet Kumar, and Suresh
  Venkatasubramanian.
\newblock The connectivity server: fast access to linkage information on the
  web.
\newblock In {\em Proceedings of 7th International World Wide Web Conference},
  pages 14--18, 1998.

\bibitem{BOLD03}
Paolo Boldi and Sebastiano Vigna.
\newblock The webgraph framework i: Compression techniques.
\newblock Technical Report 293-03, Dipartimento di Scienze dell'Informazione,
  Universit\`a di Milano, 2003.

\bibitem{BORG00}
Jose Borges.
\newblock {\em A Data Mining Model to Capture User Web Navigation Patterns}.
\newblock PhD thesis, University College London, 2000.

\bibitem{BROD00}
Andrei~Z. Broder.
\newblock Identifying and filtering near-duplicate documents.
\newblock In {\em Proceedings of the 11th Annual Symposium on Combinatorial
  Pattern Matching}, pages 1--10, 2000.

\bibitem{BUSH45}
Vannevar Bush.
\newblock As we may think.
\newblock {\em Atlantic Monthly}, 76:101--108, 1945.

\bibitem{CHEN99}
Michael Chen, Marti Hearst, Jason Hong, and James Lin.
\newblock Cha-cha: A system for organizing intranet search results.
\newblock In {\em USENIX Symposium on Internet Technologies and Systems}, 1999.

\bibitem{CHO98}
Junghoo Cho, Hector {Garcia-Molina}, and Lawrence Page.
\newblock Efficient crawling through {URL} ordering.
\newblock In {\em Proceedings of International World Wide Web Conference},
  pages 161--172, Brisbane, 1998.

\bibitem{CONK87}
Jeff Conklin.
\newblock Hypertext: An introduction and survey.
\newblock {\em IEEE Computer}, 20:17--41, 1987.

\bibitem{COOP99}
Alan Cooper.
\newblock {\em The Inmates are Running the Asylum}.
\newblock Sams, 1999.

\bibitem{DORI96}
Marco Dorigo, Vittorio Maniezzo, and Alberto Colorni.
\newblock The ant system: Optimization by a colony of cooperating agents.
\newblock {\em IEEE Transactions on Systems, Man, and Cybernetics Part B:
  Cybernetics}, 26(1):29--41, 1996.

\bibitem{FURU97}
Richard Furuta, Frank~M. {Shipman III}, Catherine~C. Marshall, Donald Brenner,
  and {Hao-wei} Hsieh.
\newblock Hypertext paths and the world-wide web: Experiences with walden's
  paths.
\newblock In {\em Proceedings of the Eighth ACM Conference on Hypertext}, pages
  167--176, Southampton, U.K., April 1997.

\bibitem{GUIN92}
Catherine Guinan and Alan~F. Smeaton.
\newblock Information retrieval from hypertext using dynamically planned guided
  tours.
\newblock In {\em Proceedings of the ACM European Conference on Hypertext},
  1992.

\bibitem{HAVE99}
Taher~H. Haveliwala.
\newblock Efficient computation of pagerank.
\newblock Technical report, Stanford University, 1999.

\bibitem{JOAC97}
Thorsten Joachims, Dayne Freitag, and Tom Mitchell.
\newblock Web{W}atcher: A tour guide for the {W}orld {W}ide {W}eb.
\newblock In {\em Proceedings of International Joint Conference on Artificial
  Intelligence}, pages 770--775, Nagoya, Japan, 1997.

\bibitem{KAMV03}
Sepandar~D. Kamvar, Taher~H. Haveliwala, Christopher~D. Manning, and Gene~H.
  Golub.
\newblock Extrapolation methods for accelerating pagerank computations.
\newblock In {\em Proceedings of the World Wide Web Conference}, Budapest,
  2003.

\bibitem{KLEI98}
Jon~M. Kleinberg.
\newblock Authoritative sources in a hyperlinked environment.
\newblock In {\em Proceedings of ACM-SIAM Symposium on Discrete Algorithms},
  pages 668--677, San Francisco, 1998.

\bibitem{KOST94}
Martijn Koster.
\newblock Robot exclusion, June 1994.

\bibitem{LEMP00}
Ronny Lempel and Shlomo Moran.
\newblock The stochastic approach for link-structure analysis ({SALSA}) and the
  {TKC} effect.
\newblock In {\em Proceedings of the 9th International World Wide Web
  Conference}, pages 387--402, 2000.

\bibitem{LEVE02a}
Mark Levene and George Loizou.
\newblock Web interaction and the navigation problem in hypertext.
\newblock In A.~Kent, J.G. Williams, and C.M. Hall, editors, {\em Encyclopedia
  of Microcomputers}, pages 381--398. Marcel Dekker, New York, NY, 2002.

\bibitem{LEVE01a}
Mark Levene and Richard Wheeldon.
\newblock A {W}eb site navigation engine.
\newblock In {\em Poster Proceedings of International World Wide Web
  Conference}, Hong Kong, 2001.

\bibitem{LEVE03}
Mark Levene and Richard Wheeldon.
\newblock Navigating the world-wide-web.
\newblock In Mark Levene and Alex Poulovassilis, editors, {\em Web Dynamics}.
  Springer-Verlag, 2003.

\bibitem{LI01}
Wen-Syan Li, K.~Selcuk Candan, Quoc Vu, and Divyakant Agrawal.
\newblock Retrieving and organizing web pages by "information unit".
\newblock In {\em Proceedings of International World Wide Web Conference},
  pages 230--244, Hong Kong, 2001.

\bibitem{MATH01}
Mazlita Mat-Hassan and Mark Levene.
\newblock Can navigational assistance improve search experience: {A} user
  study.
\newblock {\em First Monday}, 6(9), 2001.

\bibitem{NIEL90}
Jakob Nielsen.
\newblock {\em Hypertext and Hypermedia}.
\newblock Academic Press, Boston, Ma., 1990.

\bibitem{NIEL97}
Jakob Nielsen.
\newblock Search and you may find, 1997.
\newblock useit.com alertbox.

\bibitem{PAGE98}
Lawrence Page, Sergey Brin, Rajeev Motwani, and Terry Winograd.
\newblock The pagerank citation ranking: {B}ringing order to the web.
\newblock Working paper, Department of Computer Science, Stanford University,
  1998.

\bibitem{PINK00}
Brian Pinkerton.
\newblock {\em WebCrawler: Finding What People Want}.
\newblock PhD thesis, University of Washington, 2002.

\bibitem{RAND01}
Keith~H. Randall, Raymie Stata, Rajiv Wickremesinghe, and Janet~L. Wiener.
\newblock The link database: Fast access to graphs of the web.
\newblock In {\em Proceedings of the Data Compression Conference}, Snao Bird,
  Utah, April 2002.

\bibitem{REIC99}
Siegfried Reich, Leslie Carr, David~De Roure, and Wendy Hall.
\newblock Where have you been from here? : Trails in hypertext systems.
\newblock {\em ACM Computing Surveys}, 31(4), December 1999.

\bibitem{SALT98}
Gerard Salton and Chris Buckley.
\newblock Term weighting approaches in automatic text retrieval.
\newblock {\em Information Processing and Management}, 24:513--523, 1998.

\bibitem{SHIV99}
Narayanan Shivakumar and Hector Garcia-Molina.
\newblock Finding near-replicas of documents on the web.
\newblock In {\em {WEBDB}: International Workshop on the World Wide Web and
  Databases, WebDB}. LNCS, 1999.

\bibitem{SILL90}
T.J. Sillitoe, B.~Nick Rossiter, and Michael~A. Heather.
\newblock Trail management in hypertext: Database support for navigation
  through textual complex objects.
\newblock In A.~Brown and P.~Hitchcock, editors, {\em Proceedings of the 8th
  British National Conference on Databases}, pages 224--242, 1990.

\bibitem{SILV99}
Craig Silverstein, Monika Henzinger, Hannes Marais, and Michael Moricz.
\newblock Analysis of a very large altavista query log.
\newblock {\em SIGIR Forum}, 33(1):6--12, 1999.

\bibitem{SLEE01}
{Sleepycat Software}.
\newblock {\em Berkeley DB}.
\newblock New Riders Publishing, 2001.

\bibitem{WHEE03d}
Richard Wheeldon and Steve Counsell.
\newblock Making refactoring decisions in large-scale java systems: an
  empirical stance.
\newblock {\em Computing Research Repository}, cs.SE/0306098, June 2003.

\bibitem{WHEE03c}
Richard Wheeldon and Steve Counsell.
\newblock Power law distributions in class relationships.
\newblock In {\em Proceedings of 3rd International Workshop on Source Code
  Analysis and Manipulation (SCAM)}, Amsterdam, September 2003.

\bibitem{WHEE02}
Richard Wheeldon, Mark Levene, and Nadav Zin.
\newblock Autodoc: A search and navigation tool for web-based program
  documentation.
\newblock In {\em Poster Proceedings of International World Wide Web
  Conference}, Honolulu, HI, 2002.

\end{thebibliography}

\end{document}